# High cadence measurement of neutral sodium and potassium absorption during the 2009-2011 eclipse of epsilon Aurigae

R Leadbeater[1], C Buil[2], T Garrel[3], S Gorodenski[4], T Hansen[5], L Schanne[6], R Stencel[7], B Stober[8], O Thizy[9]

[1] Three Hills Observatory, The Birches, CA7 1JF, UK
[2] Castanet Tolosan Observatory, 6 Place Clemence Isaure, 31320 Castanet Tolosan, France
[3] Observatoire de Foncaude, Juvignac, France
[4] 9440 E. Newtown Ave., Dewey, Arizona, USA, 86327
[5] Reichau 216 D-87737 Boos Germany
[6] Observatory for Stellar Spectroscopy Völklingen, Hohlstrasse 19, 66333 Völklingen,German
[7] Department of Physics and Astronomy, University of Denver, 2112 East Wesley Avenue, Denver, Colorado 80208, USA
[8] Nelkenweg 14, 66791 Glan-Münchweiler, Germany
[9] Shelyak Instruments, Les Roussets, 38420 Revel, France

**Abstract**

The results of a spectroscopic survey of epsilon Aurigae during eclipse using a network of small telescopes are presented. The spectra have a resolution of 0.35 to 0.65Å and cover the period 2008 to 2012 with a typical interval of 4 days during eclipse. This paper specifically covers variations in the K I 7699Å, Na D and Mg II 4481Å lines. Absorption started increasing in the KI 7699Å line 3 months before the eclipse began in optical photometry and had not returned to pre-eclipse levels by the end of the survey March 2012, 7 months after the broadband brightness had returned to normal outside eclipse levels. The contribution of the eclipsing object to the KI 7699Å line has been isolated and shows the excess absorption increasing and decreasing in a series of steps during eclipse ingress and egress. This is interpreted as an indication of structure within the eclipsing object. The F star is totally obscured by the eclipsing object at the Na D wavelength during eclipse. The radial velocity of the F star and the mean and maximum radial velocity of the eclipsing material in front of the F star at any given time have been isolated and tracked throughout the eclipse. The quasi-periodic variations seen in the F star RV outside eclipse continued during the eclipse. It is hoped that these results can be used to constrain proposed models of the system and its components.

## 1. Background

Epsilon Aurigae is a naked eye eclipsing binary system with a period of 27.1 years and a primary eclipse of about two years. Despite having been studied for almost two centuries, our understanding of the exact nature of the system, the primary star and its largely unseen companion, is still incomplete.

At each eclipse, a new generation of astronomers equipped with the latest technology tackles the problem. A recent multiwavelength study of the system proposed that the spectral class F primary star is most likely an evolved post-AGB star and that a B5V class star is embedded in the cool (~1000 K) opaque material which causes the eclipse

*(Hoard, Howell and Stencel 2010, hereafter referred to as the HHS model).*
Interferometric measurements made during the 2009-2011 eclipse have established that the eclipsing object is an elongated opaque cigar shaped object, most likely a disc seen almost edge on, which covers the southern half of the star during eclipse *(Kloppenborg et al. 2010).*

Advances in sensor technology and the availability of affordable high resolution spectrographs allowed a network of advanced amateurs using small aperture telescopes to contribute during the 2009-2011 eclipse. The objective was to study the evolution of the optical spectrum of the system throughout the eclipse with improved time resolution compared with that achieved during previous eclipses. Over 800 spectra were collected and these are available online at
http://www.threehillsobservatory.co.uk/epsaur_spectra.htm

**2. Observations**

This paper specifically covers:

- 275 observations of the neutral potassium line at 7699 Å (hereafter referred to as K 7699) at a mean interval of 4.3 days throughout the eclipse.

- 199 observations of the neutral sodium D lines at 5890/5896 Å (hereafter referred to as Na D) at a mean interval of 4 days during the eclipse, excluding the periods around solar conjunction.

- 137 observations of the singly ionised magnesium line at 4481 Å (hereafter referred to as Mg 4481).

The observations are summarised in Table 1. The K 7699 line observations were made by Leadbeater and Schanne at a resolution of 0.35 Å using Lhires III Littrow spectrographs, modified to reach this wavelength. Various spectrograph designs with resolutions from 0.35-0.65 Å were used for the Na D line observations. The Mg 4481 line measurements were made with the eShel echelle spectrographs of Buil, Thizy and Garrel at typically 0.5 Å resolution.

**2.1** Choice of lines

Ouside eclipse, the eps Aur spectrum shows an interstellar K 7699 line *(Welty and Hobbs 2001)*. There is no detectable contribution from the F star spectrum. This was confirmed by examining spectra outside eclipse *(Lambert and Sawyer 1986, Welty and Hobbs 2001)* recorded at different phases which showed no variation in radial velocity (RV) above the level of the measurement uncertainty. If a stellar component was present this would be detectable due to the change in RV of that component. After removal of the constant interstellar component, the remaining K 7699 line uniquely describes the absorption due to the eclipsing component. Spectra of this line had also been taken during the previous eclipse, though at less frequent intervals *(Lambert and Sawyer 1986).*

The Na D lines also show strong additional absorption during eclipse *(Barsony et al 1986)* ; however, the analysis of these lines is more complex due to the presence of components from both the F star and the interstellar medium.

The Mg 4481 line arises from the F star photosphere and is absent from the cool eclipsing object spectrum due to the high excitation level of the former *(Ferluga and Hack 1985)*, so acts as a reference for changes in the F star RV.

**2.2** Reduction of spectra

The spectra were initially individually reduced by each observer. The software used is listed in Table 1. The spectra were then further reduced and line parameters measured by RL using Visual Spec software.

The pre-reduced spectra for each line of interest were first normalised using a fit to the local continuum. To maximise the global precision of the Na D and K 7699 line wavelength calibration, the original calibrations were checked using telluric lines visible in the spectra and small offsets applied as required. The estimated residual uncertainty in wavelength is 0.02A. There are no tellurics in the Mg 4481 line region so the observers' original ThAr lamp calibrations were used for this line. The telluric lines were then removed, by dividing by hot line free star spectra taken with the same equipment in the case of the K7699 line spectra or by scaling a standard telluric line template in the case of the Na D lines. Finally all spectra were corrected to the heliocentric velocity reference frame.

**3. Analysis**

For this paper, eclipse start and end dates of JD 2455070 and 2455800 have been adopted based on V-band photometric data submitted to AAVSO *(Mauclaire et al 2012)*. The parameters used for calculating the properties of the system are listed in Table 2.

**3.1**    KI 7699Å line

Figure 1 shows the evolution of the K 7699 line throughout the eclipse. Each row represents an interval of 2 days. The dates of the spectroscopic measurements are marked on the time axis. Intermediate rows between measured spectra are interpolated. Note that the interstellar component (measured from pre eclipse spectra) has been removed so the graphic shows just the contribution from the eclipsing object. Figures 2 and 3 show typical profiles for the K 7699 line with the interstellar component present and removed, respectively.

Figure 4 shows the variation in the strength (equivalent width, EW) of the K 7699 line including the interstellar component, with measurements from the previous eclipse *(Lambert and Sawyer1986)* superimposed for comparison. Figure 5 shows the same EWs for the K 7699 line with the interstellar component removed. The estimated uncertainty in EW is 15 mÅ based on repeat measurements and comparison with coincident observations made by other observers during eclipse ingress *(Ketzeback 2009)*.

The RV trend for the K 7699 line after removal of the interstellar component is shown in Figure 6. The maximum velocity component in the material in front of the F star at any given time during ingress (line profile red edge) and egress (line profile blue edge) is also plotted.

**3.2** Na D lines

Figure 7 shows the evolution of the Na D lines throughout the eclipse. Each row represents an interval of 2 days. To produce a consistent set of spectra, higher resolution spectra were filtered to give a common resolution of 0.65 Å. The dates of the spectroscopic measurements are marked on the time axis. Intermediate rows between measured spectra are interpolated. Figure 8 shows typical $D_2$ line profiles throughout the eclipse, while Figure 9 shows a selection of full Na D line profiles at a higher resolution of 0.35 Å obtained during the second half of the eclipse. The total EW of the Na D line (sum of $D_1$ and $D_2$) as a function of time before and during the eclipse is plotted in Figure 10.

The RVs of the Na D lines (mean of $D_1$ and $D_2$) and of the Mg 4481 line as a function of time before and during the eclipse are plotted in Figure 11. Also shown is a linear fit to the Mg 4481 line data and, for comparison, the radial component of the orbital velocity of the F star *(Stefanik et al 2010)*.

**4. Discussion**

**4.1.** The extent of the eclipsing object

Interferometric imaging has shown the drop in brightness during eclipse to be due to an elongated opaque object (a thin rotating disc seen almost edge on) which covers the southern half of the F star *(Kloppenborg et al 2010)*. The changes we see in the spectrum during eclipse arise from absorption within a gaseous region surrounding this opaque object. Increased absorption was first detected in the K 7699 line 95 days before photometric first contact and was still detectable at the end of the survey, 215 days after 4[th] contact. Adopting a scale of 3.8 AU for the radius of the disc as per the HHS model, we estimate that this gaseous region extends beyond the outer edges of the opaque region by 1.2 AU on the ingress side and at least 2.6 AU on the egress side.

Outside eclipse the Na D lines, a combination of contributions from the F star and the interstellar medium, are not fully saturated. During eclipse, however, the lines saturate and the residual flux in the core of the lines falls close to zero. This is seen particularly clearly in the higher resolution spectra shown in Figure 9 taken during the second half of the eclipse , in which the eclipsing disc component of the line, blue-shifted during this half of the eclipse, shows a distinctly flattened bottom at just 0.03 of the flux relative to the continuum. Given that the opaque region eclipses only the southern half of the F star (as seen in the interferometric images) this suggests that the Na absorbing region extends at least the F star radius (0.6 AU) above the opaque region, completely covering the F star.

A small peak is visible, ~10 km/s bluewards of the rest wavelength, in the high resolution Na D absorption lines during the second half of the eclipse *(Figure 9)*. This was also observed during the previous eclipse *(Barsony et al 1986)*. It is likely that this is due to a partial separation of the eclipsing disc/F star components (both blue-shifted during this half of the eclipse) and the interstellar component. The same effect is seen in the K 7699 line where the separation of the two components is clear *(Figure 2)*.

The EW of the Na D lines at the end of the campaign period was still higher than pre-eclipse values (2.2 Å vs. 1.9 Å) ; however, this could be due at least in part to the relative separation of the components. (Because the line is saturated, the EW will depend on the overlap between the components which, in turn, depends on the relative RV). Extended post-eclipse monitoring could help clarify this.

The EW of both the K 7699 and Na D lines dropped around mid-eclipse. In the case of the saturated Na D line this could be caused, for example, by a narrowing of the already saturated line profile rather than a net reduction in the amount of absorbing material. This would not be the case for the unsaturated K 7699 line, however, and suggests that either there is less absorbing material in these inner regions of the disc or that the conditions closer to the central star do not allow the particular transition (due to radiation from the central star, for example). The interval between the leading and trailing EW maxima for the K 7699 line is 265 days, which corresponds to a region 2.1 AU in diameter.

There is significant asymmetry in the K 7699 line excess absorption between the leading and trailing regions of the disc. The trailing maximum is 70% higher than the leading maximum (780 mÅ vs. 460 mÅ, see Figure 5). As already mentioned, the tail of the absorption extends significantly further on the egress side. The lateral extent of the main region of absorption, however, is similar for both halves (200 days from 30% to maximum absorption during ingress compared with 220 days during egress).

The minimum flux in the K 7699 line profile remained significantly above zero during the eclipse (the minimum level was 0.18 relative to the continuum) so provided that the material producing this line extends above the disc to the same extent as that for the Na D line, covering the F star completely, we can conclude that when we look at the K 7699 line we are seeing through the full thickness of the material and, therefore, the line profile includes contributions from all depths within the absorbing region in front of the F star at the time.

**4.2** Comparison with the previous eclipse

There is good overall agreement between the total K 7699 line EW trend during this eclipse and measurements made by Lambert and Sawyer on this line during the previous eclipse, offset by 9896 days. The largest differences occur at RJD 55340 and 55615 but Lambert and Sawyer plotted a smooth curve through their more sparse data and attributed the residual scatter to measurement error so it is not clear if these differences are significant.

**4.3  Structure within the disc**

The trend of excess K 7699 absorption during eclipse *(Figure 5)* did not progress smoothly during ingress and egress but proceeded in a series of steps.  The steps during ingress have been interpreted as an indication of structure within the disc material *(Leadbeater and Stencel 2010)*. There is no obvious symmetry in these features between ingress and egress, as might be expected for a simple system of concentric circular rings, though more complex structures such as an elliptical system or spiral density waves produced by the tidal influence of the F star, as seen in other circumstellar discs (e.g., *Muto et al 2012),* are not ruled out.

(Note that features are also seen in the Na D line total EW trend but these do not correspond with the steps seen in the K 7699 EW and may be caused by interactions between the various components which make up the line. These interactions are discussed in Section 4.5.)

**4.4  The disc rotation**

The trend of the K 7699 line  RV after removal of the interstellar component *(Figure 6)* is smooth throughout the eclipse with little scatter (except at the start and end of eclipse when the line is weak) and no obvious short timescale features. Given the known quasi-periodic variability seen in the RV of the F star lines *(Stefanik et al 2010)*, this is additional confirmation that the K 7699 line is not significantly contaminated by any contribution from the F star.  Since we are seeing through the full depth of the disc at this wavelength, the  RV values plotted here are a measure of the mean velocity along our line of sight of all the vectors of the orbiting components of the absorbing material in front of the F star at a particular phase of the eclipse. While it may prove possible to model such a parameter given a detailed description of the properties and distribution of the material surrounding the disc, a perhaps simpler parameter, the maximum velocity component in the line, has also been plotted in Figure 6.  This is somewhat more difficult to measure as it involves estimating the wavelength of the edge of the line where it meets the continuum, and this is reflected in the increased scatter.  It has the potential, however, to be used to calculate an orbital velocity curve for the disc since, for material in Keplerian orbits, the maximum radial velocity at a given time will be due to the innermost material in front of the F star at that time viewed tangentially and, hence, gives a direct measurement of the orbital speed of that material.  This has already been attempted using the data from the first half of the eclipse *(Leadbeater and Stencel 2010)* where a figure of 5.3 solar masses was estimated for the disc component. However, this value is sensitive to the RV adopted for the eclipsing component as a whole due to its orbital motion, which is not known currently. We speculate that this orbital motion is, at least in part, responsible for the asymmetry seen in the RV curve from ingress to egress, (higher values of RV seen on egress and a declining RV in the later part of the eclipse compared with the level RV during ingress) ; however, a full orbital solution for the system will be needed to clarify this.

**4.5** The radial velocity of the Na D and Mg 4481 lines

The Na D lines are a combination of components from the eclipsing disc, the interstellar medium and the F star. The interstellar component will be constant as for the K 7699 line but the F star RV will have an orbital component and quasi-periodic variations *(Stefanik et al 2010)*. The net result *(Figure 11)*, although broadly showing the same swing from red to blue during eclipse, is quite different in detail from that of the disc component extracted from the K 7699 line.

Not enough is known currently about the interstellar or F star components to allow the net effect of the eclipsing disc to be isolated in the Na D line. This might be possible in the future using more data obtained outside of eclipse. (Throughout the orbit, the Na D line F star component will move independently of the interstellar component so it should be possible to isolate the two components .) The Na D line also becomes saturated during the eclipse so the line profile is not a simple linear combination of the 3 components. Nonetheless, we can make some qualitative observations.

The Mg 4481 line is produced by the F star only *(Ferluga and Hack 1985)* so the RV of this line can be used as a reference for the F star Na D component RV during eclipse. The Mg 4481 line data clearly show that the quasi-periodic variations in F star RV seen outside eclipse continued during the eclipse. Note in Figure 11 that the Na D line RV follows these variations in the interval RJD 55080-55150. The disc component then starts to dominate, rapidly increasing the RV to ~ +16 km/s before reversing through mid eclipse to ~ -23 km/s. There is again some correlation between Na D and Mg 4481 RV in the interval RJD 55570-55700, but the disc absorption strongly saturates the Na D lines during this period so the sensitivity to the F star variations is expected to be low. By ~ RJD 55800 the RV had returned close to the F star RV as measured by the Mg 4481 line.

**5. Further Work**

Additional spectra taken outside eclipse are needed to determine the extent of the absorption due to the eclipsing disc on the egress side and to separate the interstellar, F star, and eclipsing disc components in the Na D lines.

The echelle spectra used here to study the Na D and Mg 4481 lines also show changes in many other lines during eclipse. A similar analysis of these lines may reveal more information about the structure and conditions within the eclipsing disc.

**6. Acknowledgements**

We are grateful to Jeff Hopkins for organizing the International Campaign for the observation of epsilon Aurigae during this eclipse and to all who submitted observations. We thank the epsilon Aurigae spectral monitoring team at Apache Point Observatory (W. Ketzeback, J.Barentine, et al.) for allowing us access to their K 7699 line data during ingress. We acknowledge with thanks the variable star observations from the AAVSO International Database contributed by observers worldwide and used in this research. R.E.S. is grateful for the bequest of William Herschel Womble to the University of Denver in support of astronomy, and for support under National Science Foundation grant # AST1016678 to the University of Denver.

**Tables**

Table 1  Summary of observations

| OBSERVER | APERTURE (mm) | SPECTROGRAPH | REDUCTION SOFTWARE | NUMBER OF OBSERVATIONS | | |
|---|---|---|---|---|---|---|
| | | | | K 7699 | Na D | Mg 4481 |
| Buil | 280 | eShel** | 5 | | 96 | 94 |
| Garrel | 280 | Lhires III* | 1,3 | | 4 | |
| | | eShel | 5 | | 24 | 24 |
| Gorodenski | 405 | Lhires III | 1,3 | | 26 | |
| Hansen | 280 | Lhires III | 1,3 | | 4 | |
| Leadbeater | 200/280 | Lhires III | 1,3 | 259 | 12 | |
| Schanne | 355 | Lhires III | 4 | 16 | 6 | |
| Stober | 300 | custom echelle | 4 | | 8 | |
| Thizy | 280 | eShel | 5 | | 19 | 19 |
| | | | | | | |
| TOTAL | | | | 275 | 199 | 137 |

    1   IRIS                     www.astrosurf.com/buil/us/iris/iris.htm
    2   ISIS                     www.astrosurf.com/buil/isis/isis.htm
    3   Visual Spec       www.astrosurf.com/vdesnoux/
    4   ESO-MIDAS     www.eso.org/sci/software/esomidas/
    5   AudeLA-ReShel  http://www.audela.org

   **  eShel              www.shelyak.com/dossier.php?id_dossier=47
   *   Lhires III        www.shelyak.com/dossier.php?id_dossier=46

Table 2  Adopted system parameters

| PARAMETER | VALUE | REFERENCE |
|---|---|---|
| disc radius | 3.8 AU | HHS 2010 |
| F star radius | 135 $R_{sun}$ | HHS 2010 |
| mean eclipse duration | 710 days | Parthasarathy and Frueh 1986 |

**Figures**

Fig 1

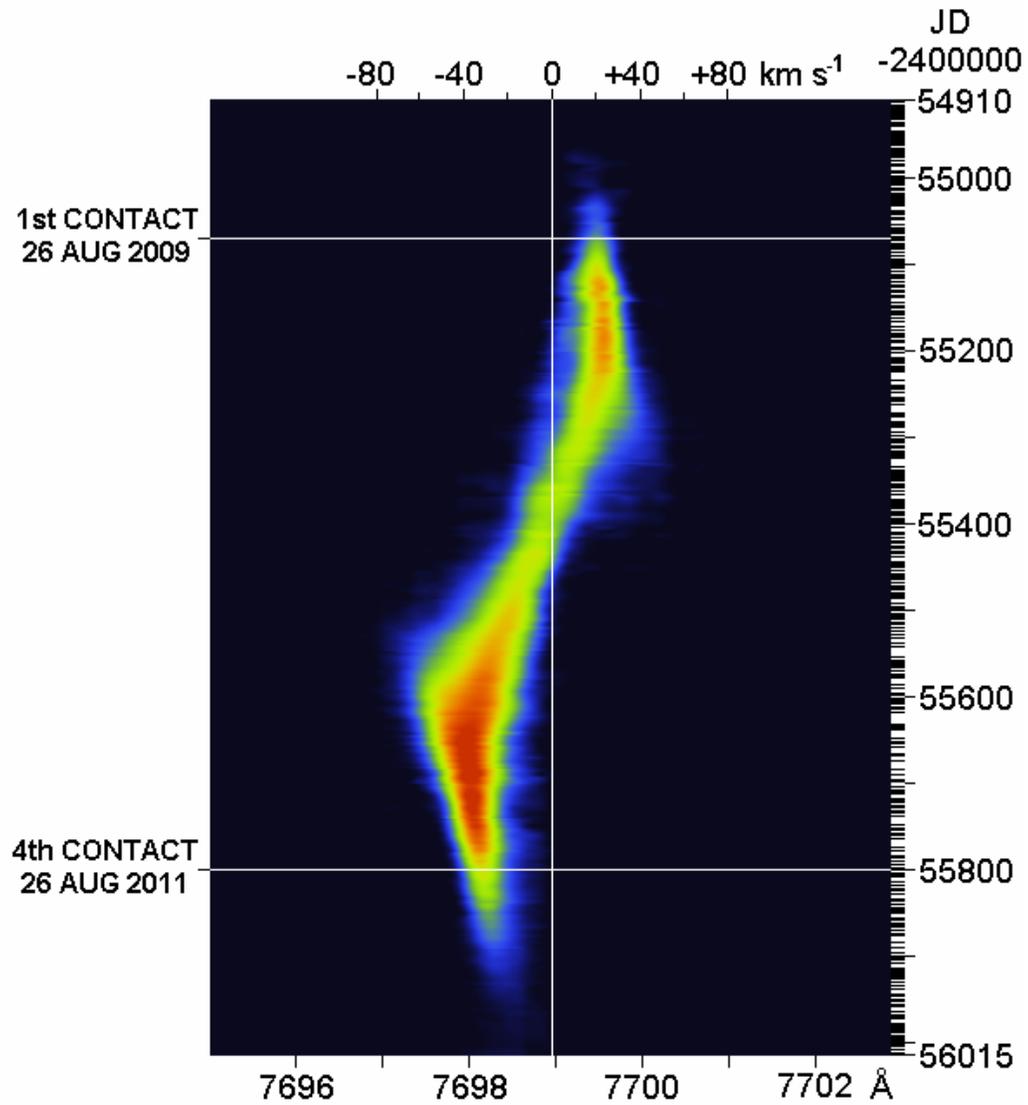

Plot showing the evolution of the K 7699 line after removal of the interstellar component (The colours indicate the degree of absorption, increasing from blue to red). Tick marks on the right-hand y-axis indicate actual measurements. Intermediate rows are interpolated.

Fig 2

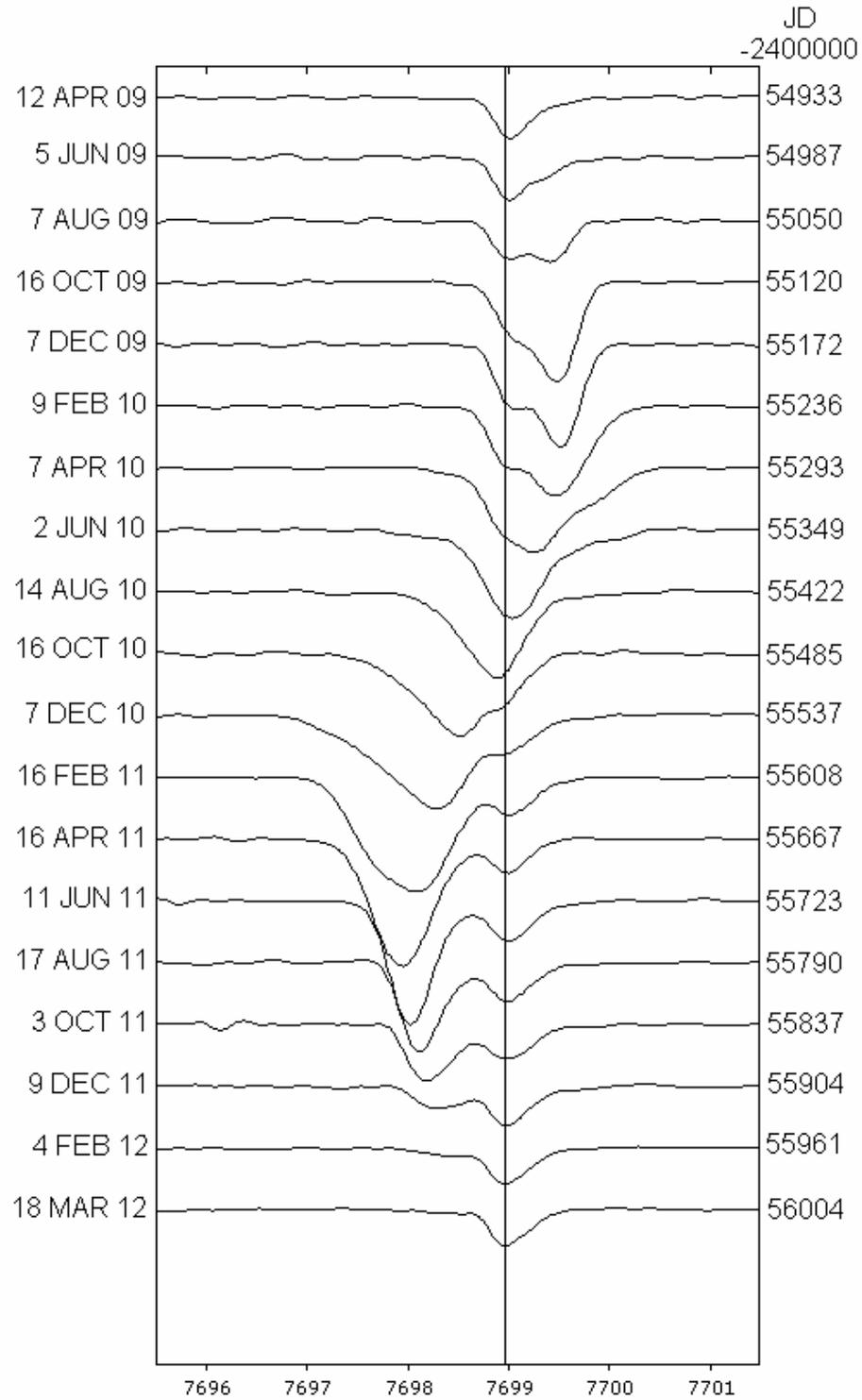

A selection of typical K 7699 line profiles with the interstellar component included.

Fig 3

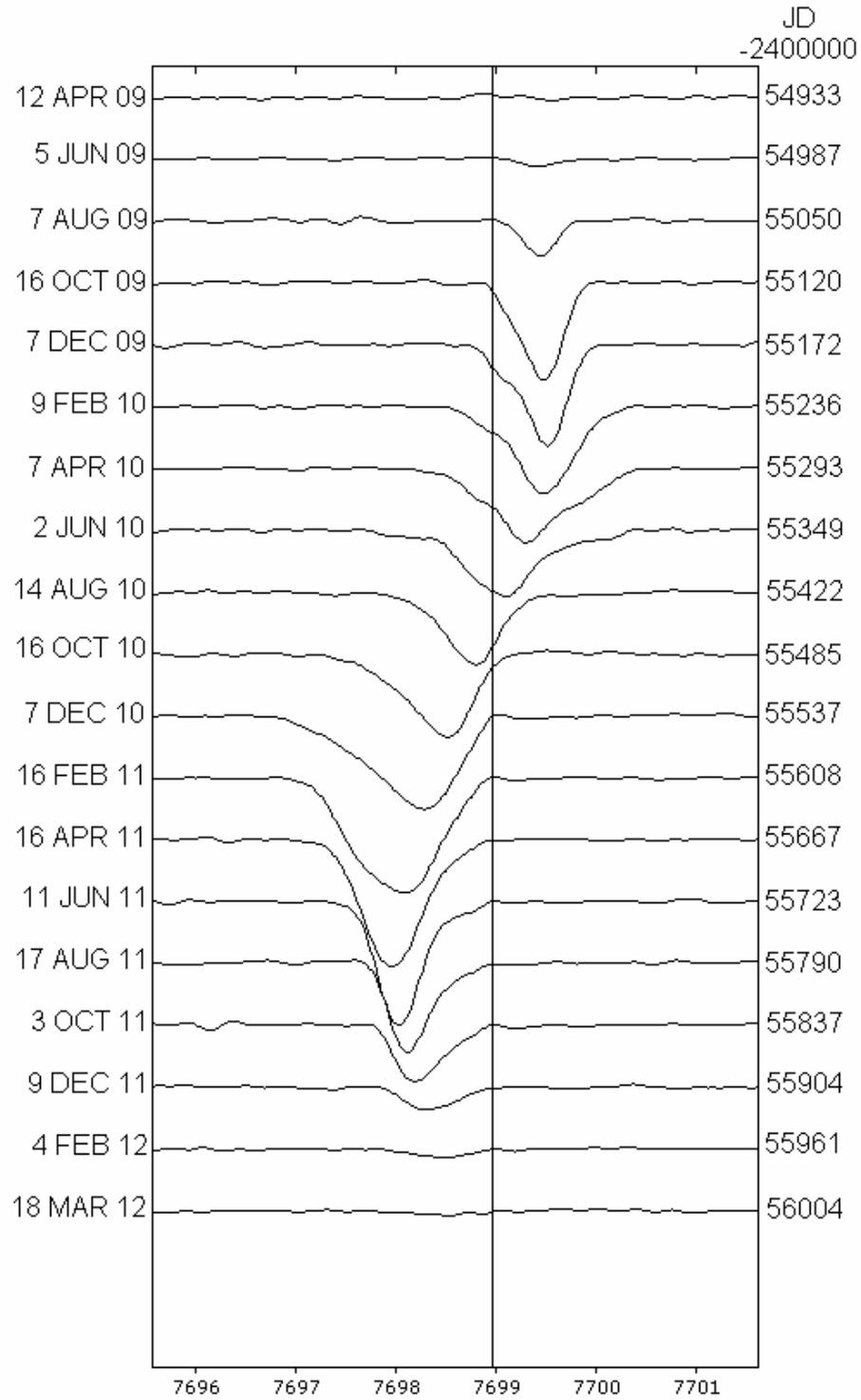

A selection of typical K 7699 line profiles with the interstellar component removed based on pre-eclipse measurements.

Fig 4

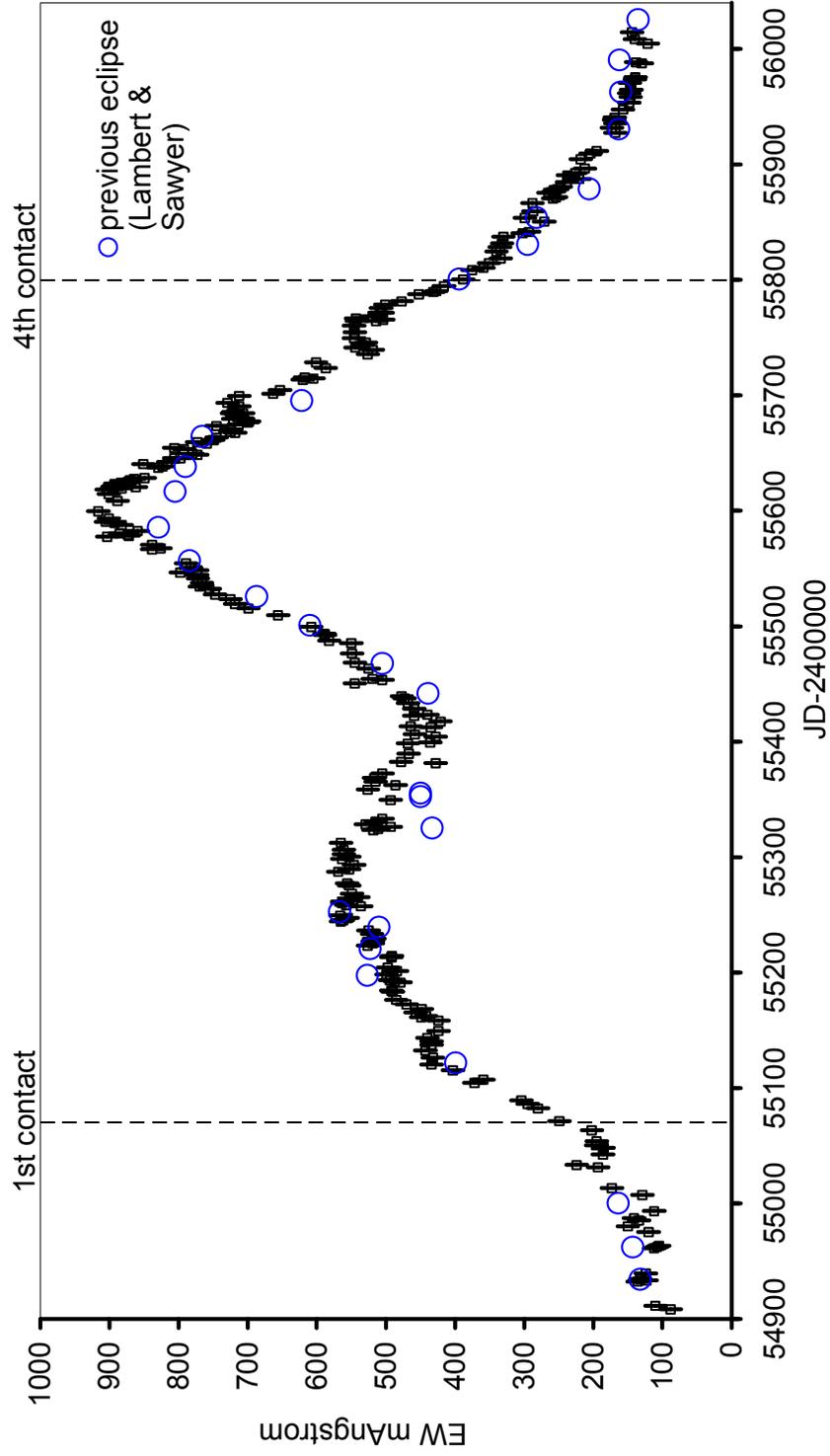

K 7699 line total equivalent width (including the interstellar component) compared with 1983 eclipse data shifted in time by +9896 days.

Fig 5

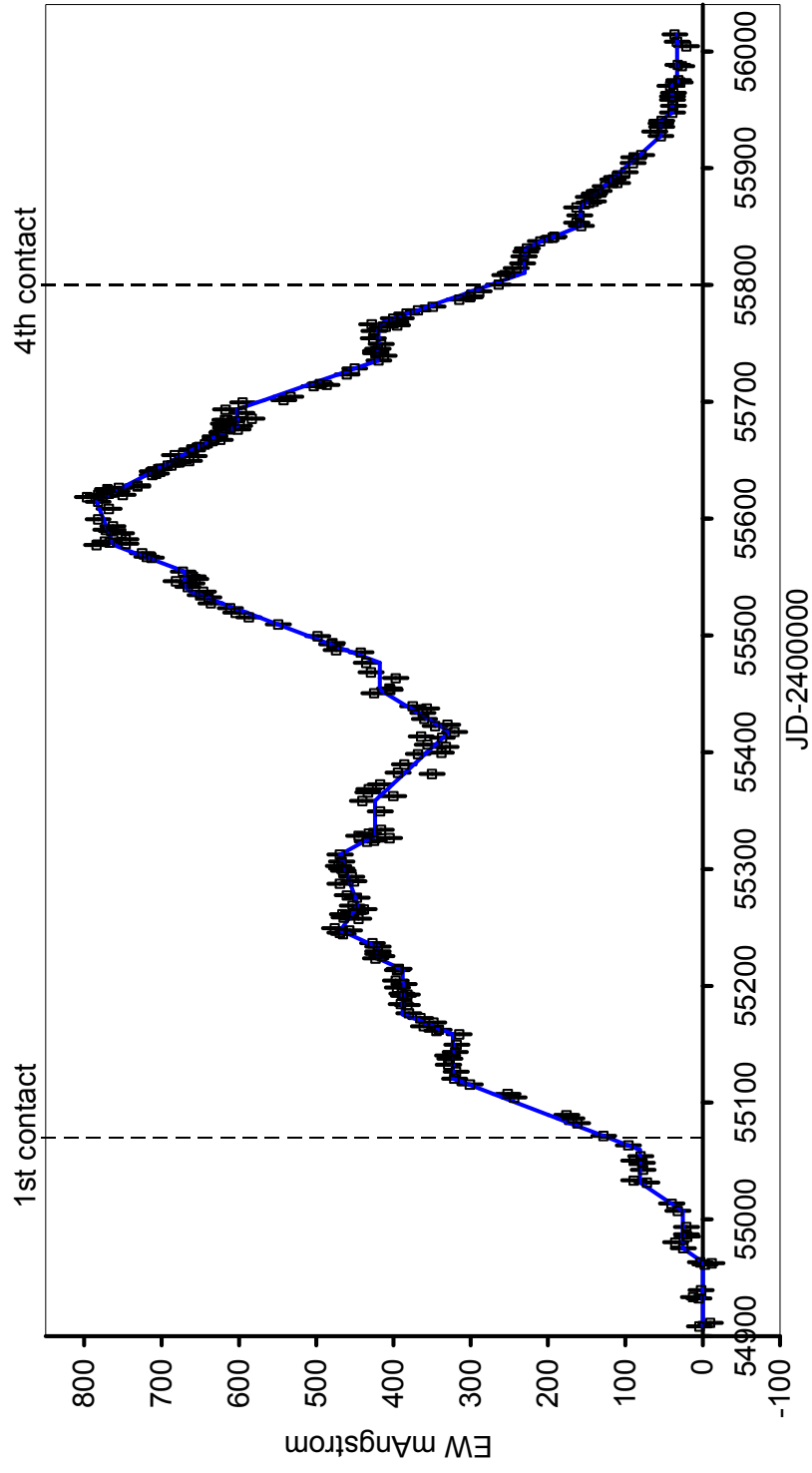

K 7699 excess equivalent width after removal of the interstellar component. The trend line indicates the stepwise progression.

Fig 6

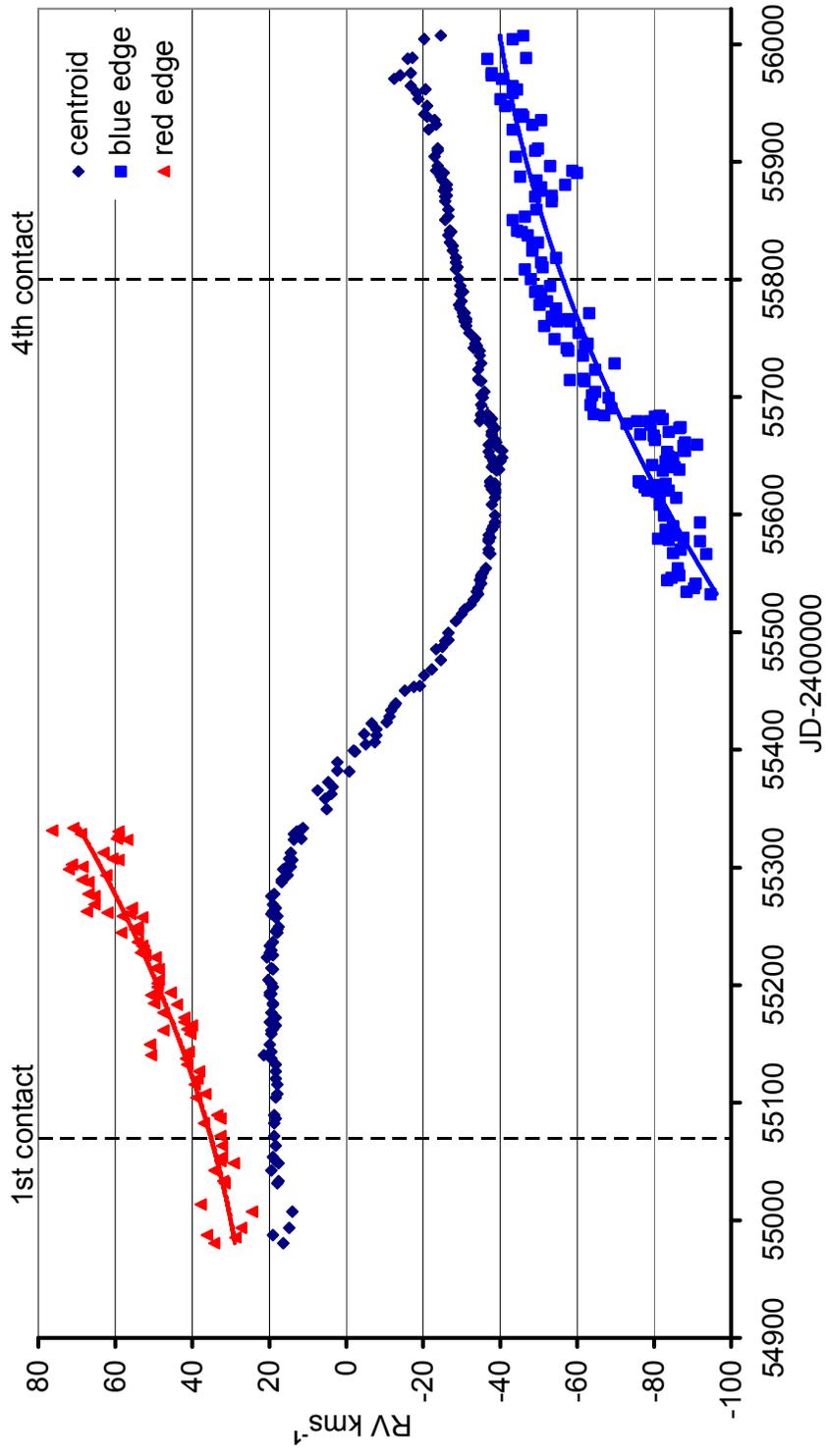

Radial velocity of the K 7699 line (heliocentric). The red and blue edge values are the maximum absolute RV in the line profile.

Fig 7

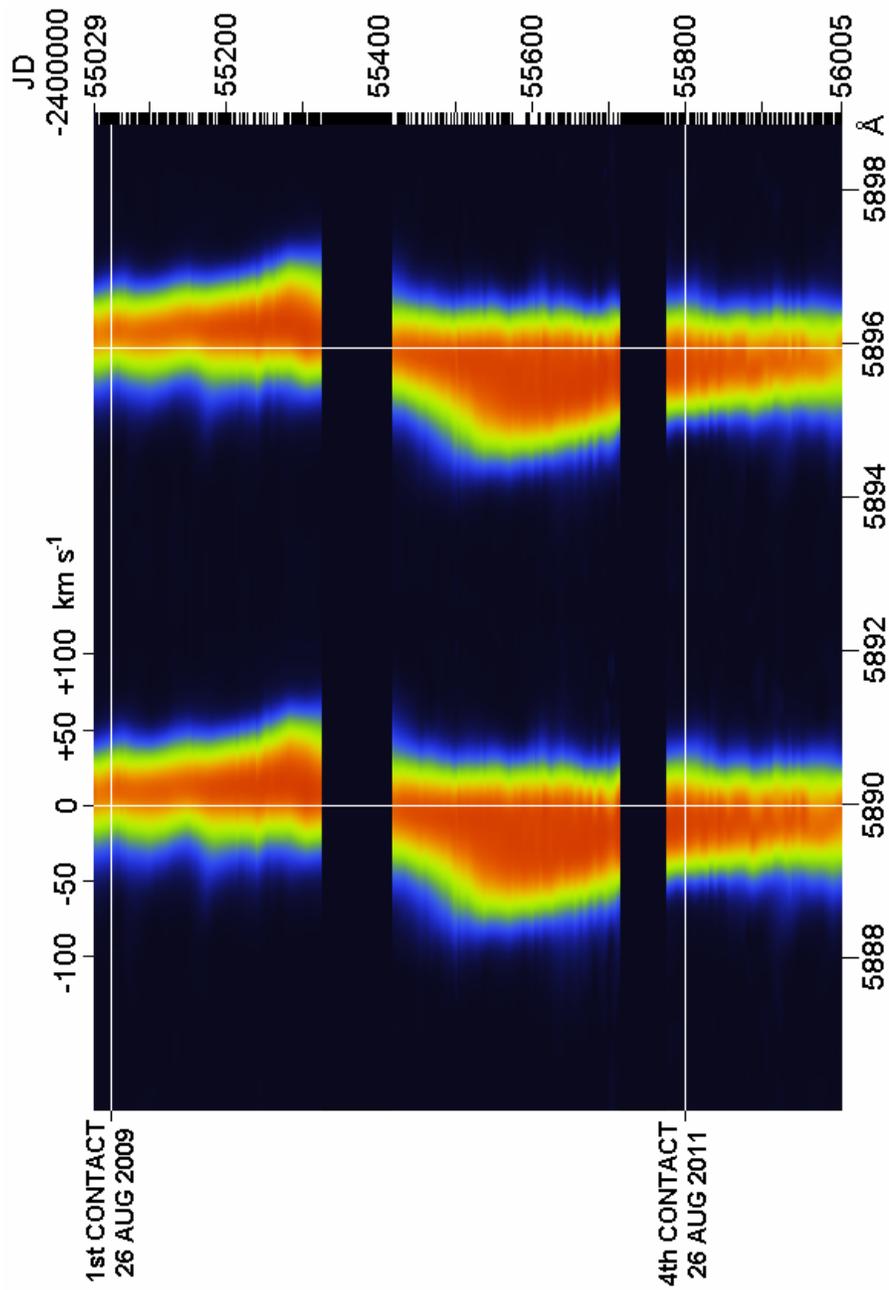

Plot showing the evolution of the Na D lines (The colours indicate the degree of absorption, increasing from blue to red). Tick marks on the right-hand y-axis indicate actual measurements. Intermediate rows are interpolated. The gaps are the periods around solar conjunction.

Fig 8

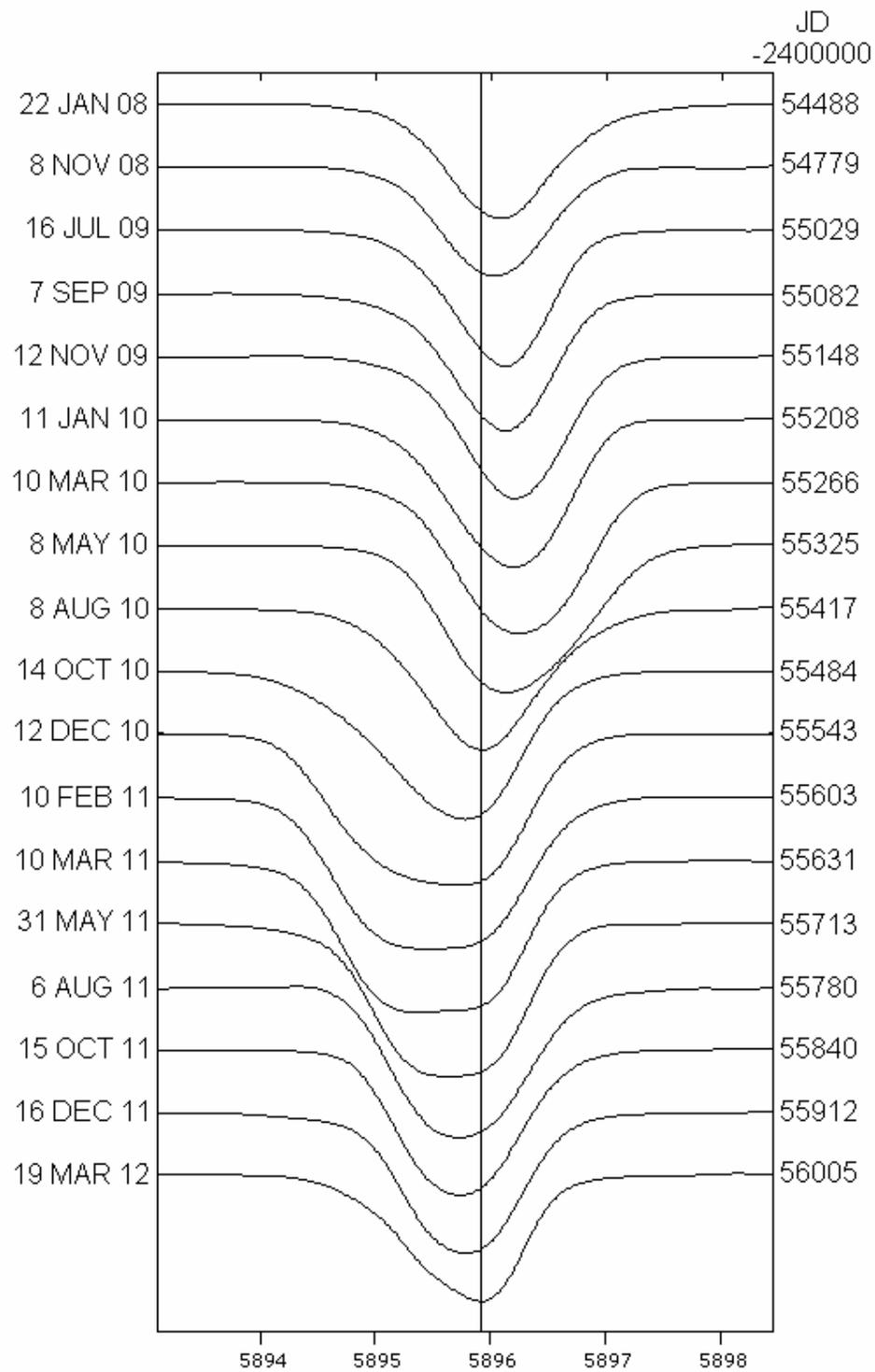

A selection of typical profiles for the Na $D_2$ line at 0.65 Å resolution.

Fig 9

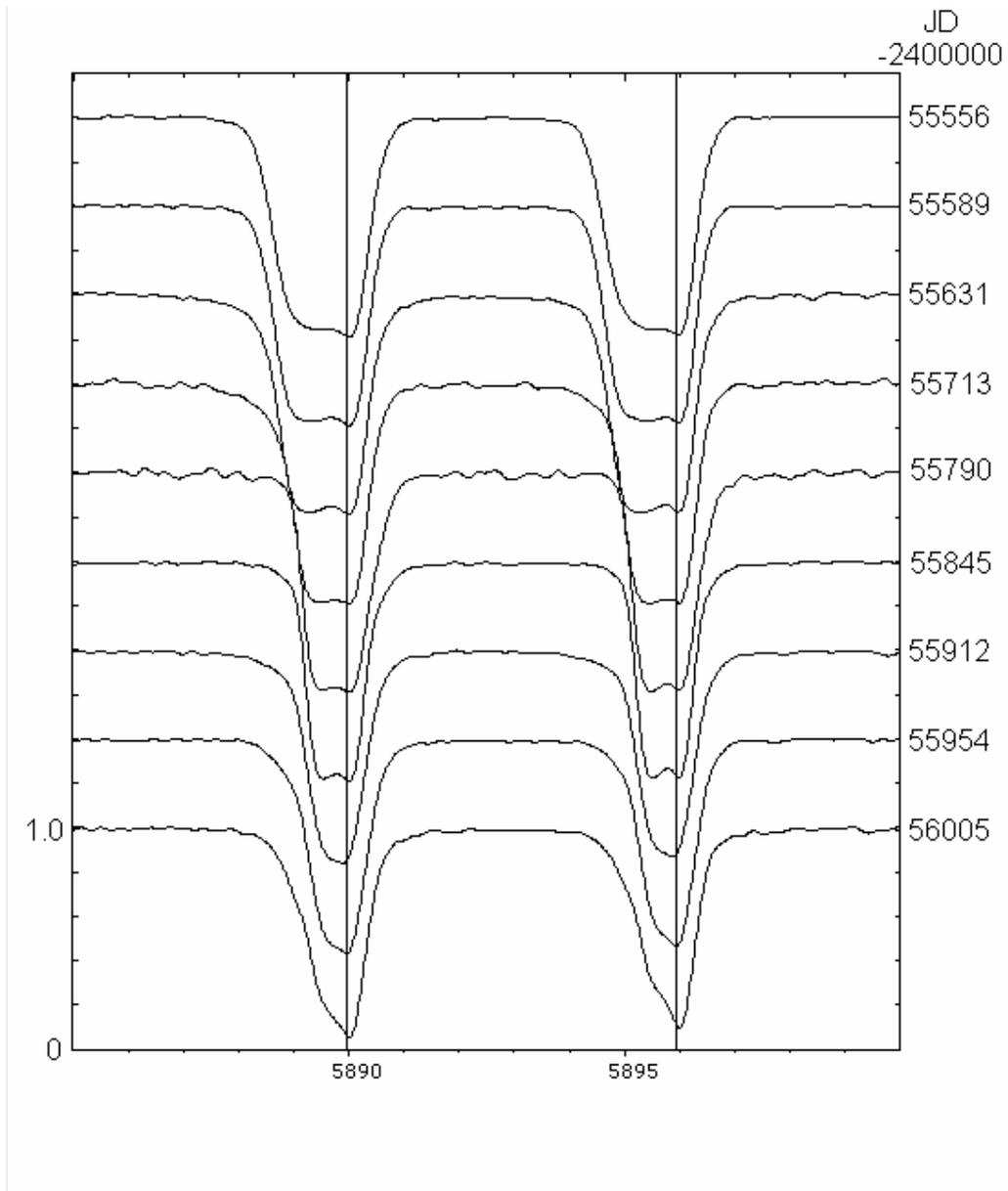

A selection of Na D line profiles covering the second half of the eclipse at 0.35 Å resolution.

Fig 10

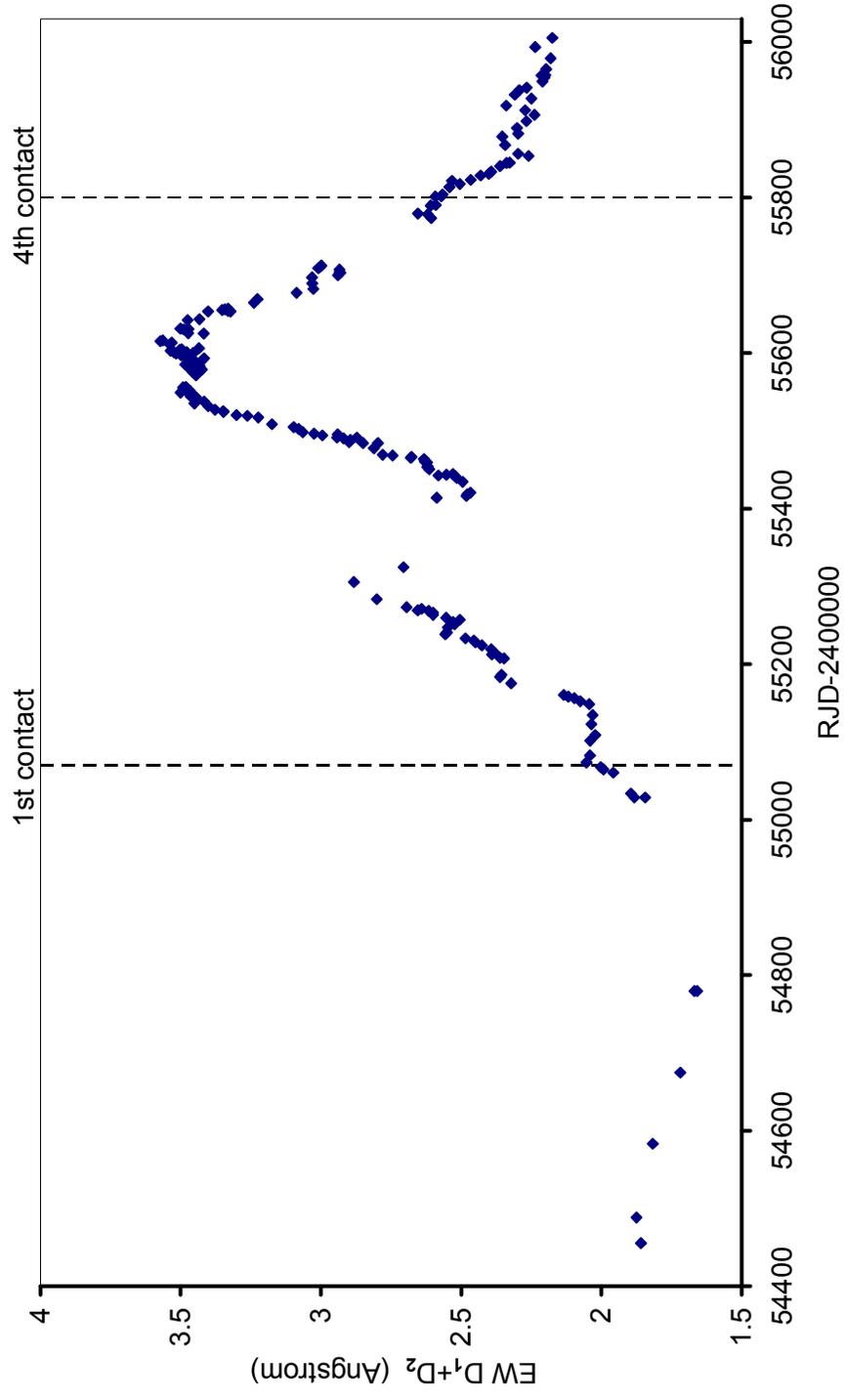

Na D line equivalent width (sum of both line components) as a function of tie before and during eclipse.

Fig 11

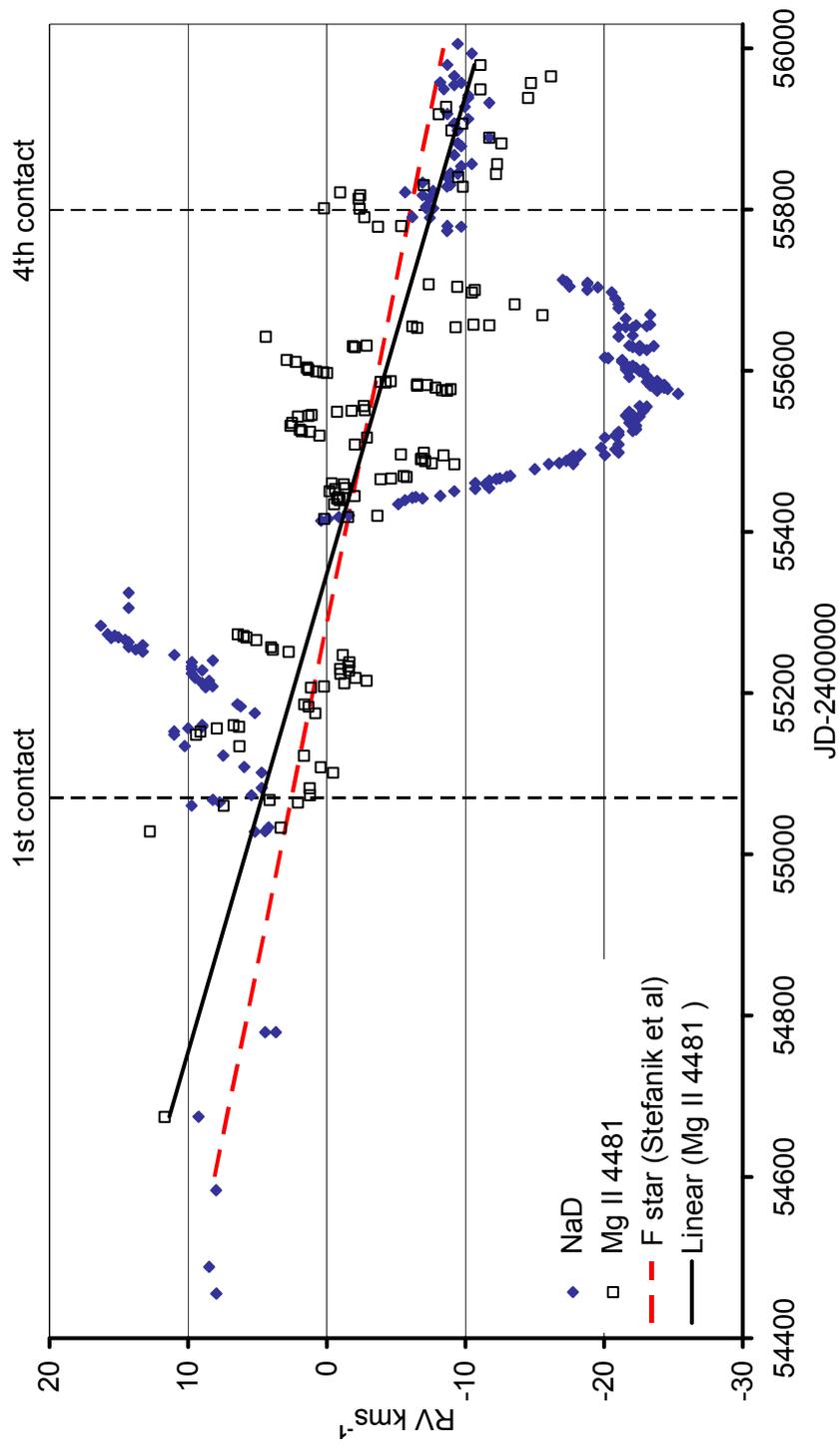

Radial velocity of the Na D lines (mean of both lines) and the Mg 4481 line. A linear fit to the Mg 4481 line data and the F star RV (based on Stefanik et al 2010) are also shown. All velocities are heliocentric.